\documentclass[a4paper,11pt]{article}
\pdfoutput=1

\usepackage{jheppub}

\usepackage{graphicx}
\usepackage{amsmath}
\usepackage{amssymb}
\usepackage{amsfonts}
\usepackage{dcolumn}
\usepackage{bm}
\usepackage[dvipsnames]{xcolor}
\usepackage[utf8]{inputenc}
\usepackage{subfigure}
\usepackage{gensymb}
\usepackage{cleveref}
\usepackage{colortbl}
\usepackage{tabularx}
\usepackage{tabulary}

\usepackage[T1]{fontenc}
\usepackage{pstricks}
\usepackage{color}
\usepackage{multirow}
\usepackage{slashed}
\usepackage{mathtools}

\usepackage[force]{feynmp-auto}
\usepackage{verbatim}

\newcommand{\Hub}{\mathcal{H}}

\newcommand{\N}{\mathcal{N}}

\DeclareMathAlphabet{\mathpzc}{OT1}{pzc}{m}{it}

\hypersetup{colorlinks,linkcolor={blue},citecolor={teal},urlcolor={violet}}

\newcommand{\SOTON}{Department of Physics and Astronomy, University of Southampton, SO17 1BJ Southampton, United Kingdom}

\begin{document}

\title{Spontaneously stabilised dark matter from a fermiophobic $U(1)'$ gauge symmetry}

\author[a,1]{B. Fu,\note{\url{https://orcid.org/0000-0003-2270-8352}}}
\author[a,2]{S. F. King\note{\url{https://orcid.org/0000-0002-4351-7507}}}

\affiliation[a]{\SOTON}

\emailAdd{B.Fu@soton.ac.uk}
\emailAdd{king@soton.ac.uk}

\abstract{
We consider the possibility that dark matter is stabilised by a discrete $Z_2$ symmetry which arises from a subgroup of a 
$U(1)'$ gauge symmetry, spontaneously broken by integer charged scalars, and 
under which the chiral quarks and leptons do not carry any charges.
A chiral fermion $\chi$ with half-integer charge is odd under the preserved $Z_2$, and hence becomes a stable dark matter candidate, being produced through couplings to right-handed neutrinos with vector-like $U(1)'$ charges, as in the type Ib seesaw mechanism. We calculate the relic abundance in such a low energy effective seesaw model containing few parameters, then 
consider a high energy renormalisable model with a  complete fourth family of vector-like fermions, where the chiral quark and lepton masses arise from a seesaw-like mechanism. With the inclusion of the fourth family, the lightest vector-like quark can contribute to the dark matter production, enlarging the allowed parameter space that we explore. 
}

\maketitle

\tableofcontents

\section{Introduction}

The origin of neutrino masses and their mixing, as evidenced by the neutrino oscillation experiments \cite{2016NuPhB.908....1O}, remains one of the most interesting open questions of physics beyond the Standard Model (SM). In the past half-century, theorists have invented hundreds of models to interpret the existence of the neutrino masses and most of them lead to an effective dimension-five Weinberg operator \cite{Weinberg:1979sa}. 
Among those models, the most popular and well-studied ones are the tree-level realisations of the Weinberg operator, namely the type I \cite{Minkowski:1977sc,Yanagida:1979as,GellMann:1980vs,Mohapatra:1979ia}, II \cite{Magg:1980ut,Schechter:1980gr,Wetterich:1981bx,Lazarides:1980nt,Mohapatra:1980yp,Ma:1998dx} and III \cite{Foot:1988aq,Ma:1998dn,Ma:2002pf,Hambye:2003rt} seesaw models. However, the difficulty in generating proper neutrino mass naturally with large seesaw couplings and small right-handed (RH) neutrino masses simultaneously reduces the experimental testability of these models, and some low scale seesaw models with extended RH neutrino sectors such as the inverse seesaw model \cite{Mohapatra:1986bd}, the linear seesaw model \cite{Akhmedov:1995ip,Malinsky:2005bi} and other radiative models \cite{Zee:1980ai,Ma:2009dk,Bonnet:2012kz,Cai:2017jrq} have been proposed to make the models more testable. 

Another great mystery unanswered by the SM is that of cosmological dark matter (DM), which is commonly thought to be some kind of massive new particle that is stable on cosmological timescales. Although many DM candidates have been proposed, the most common mechanism to account for their stability is to invent a discrete symmetry, the simplest example being $Z_2$, under which the dark matter candidate is odd, while the SM particles are even, where such models may be related to neutrino mass and mixing~\cite{Ma:2006km,Cline:2013gha,Heikinheimo:2017ofk,Becker:2018rve,Chianese:2018dsz,Bandyopadhyay:2018qcv,Chianese:2019epo,Dasgupta:2019lha,Liu:2020mxj,Chianese:2020yjo,Cheng:2020gut,Chianese:2020khl,Nam:2020byw,Biswas:2021kio,Chang:2021ose,Borah:2021pet}.
Although this approach can explain the mystery of invisible dark matter, accounting for about a quarter of the energy density of the universe \cite{Aghanim:2018eyx}, the origin of the discrete symmetry such as $Z_2$ is rarely considered in the literature, but instead is often just imposed, for example as in the case of R-parity in supersymmetry (SUSY). 
Although discrete symmetries are widely used in model building \cite{deMedeirosVarzielas:2005qg,deMedeirosVarzielas:2006fc,King:2006np,Branco:2011iw,King:2011ab,Cooper:2012wf,King:2013eh,Ding:2013bpa,King:2014nza,Karozas:2014aha,Bjorkeroth:2015ora,Altarelli:2010gt,Ma:2001dn,Babu:2002dz,Altarelli:2005yp}, 
the SM does not contain such discrete symmetries, only gauge symmetries and accidental (approximate) global symmetries. 
Consequently, there is good motivation to seek the origin of discrete symmetries as subgroups of gauge symmetries.

Recently a new version of the type I seesaw mechanism, named as the type Ib seesaw mechanism \cite{Hernandez-Garcia:2019uof}, that can be just as testable as the low scale seesaw models above has been proposed, with the light neutrino masses originating from a new type of Weinberg operator involving two Higgs doublets and a Dirac heavy neutrino. It has been shown that the model cannot only be extended to include dark matter via a neutrino portal \cite{Chianese:2021toe} but can also produce baryon asymmetry in a variant version \cite{Fu:2021fyk}. However in this model, as in many such models, 
both the type Ib seesaw model itself and the inclusion of dark matter via a neutrino portal requires additional imposed discrete symmetries whose origin is not explained.

In this paper, we consider the possibility that dark matter is stabilised by a discrete $Z_2$ symmetry which arises from a subgroup of a 
$U(1)'$ gauge symmetry, and under which the chiral quarks and leptons do not carry any charges.
A chiral fermion $\chi$ with a half-integer charge is odd under the preserved $Z_2$, and hence becomes a stable dark matter candidate, being produced through couplings to right-handed neutrinos with vector-like $U(1)'$ charges, as in the type Ib seesaw mechanism. However, in the present model, no discrete symmetries are required to be added by hand.
Indeed our proposed model is a $U(1)'$ gauge extension of the SM $SU(3)_c \times SU(2)_L \times U(1)_Y$ symmetry, where the 
$U(1)'$ is broken into a $Z_2$ symmetry spontaneously by the vacuum expectation value (VEV) of an integer charged scalar singlet, together with integer charged Higgs doublets.
In the minimal type Ib seesaw model, the light neutrino masses originate from a new type of Weinberg operator involving two Higgs doublets and a heavy Dirac neutrino constructed from the vector-like right-handed neutrinos.
Assuming the heavy Dirac neutrino is around the GeV scale, we focus on a scenario where the dark matter candidate and the new gauge boson are above TeV scale and explore the parameter space of the model providing the correct dark matter relic abundance.
However, in such a minimal effective model, chiral quark and lepton masses arise from non-renormalisable operators.
To construct a renormalisable model, we consider a complete fourth family of vector-like fermions, in which 
the chiral quark and lepton masses arise from a seesaw-like mechanism. With the inclusion of the fourth family, the lightest vector-like quark can contribute to the dark matter production, enlarging the allowed parameter space that we explore.

The paper is organised as follows. In Sec.\ref{model_eff}, we start with the extension of the minimal type Ib seesaw model and discuss the allowed parameter space in the model assuming the correct dark matter relic abundance. We also derive the required sensitivity of direct and indirect dark matter detections to find the dark fermion. In Sec.\ref{sec:four} we show how the model is completed with a fourth generation of vector-like fermions and recompute the allowed parameter space. Finally, we summarise and conclude in Sec.\ref{sec:Concl}.

\section{Extension of the minimal type Ib seesaw model as an effective model \label{model_eff}}

\begin{table}[t!]
\centering
\begin{tabular}{|c|c|c|c|c|c|c|c|c|c|c|c|}
\hline 
& ${q_L}_\alpha$ & ${u_R}_\beta$ & ${d_R}_\beta$ & ${\ell_L}_\alpha$ & ${e_R}_\beta$ & $\Phi_{1}$ & $\Phi_{2}$ & $N_{\mathrm{R}1}$ & $N_{\mathrm{R}2}$ & $\chi_{L,R}$ & $\phi$\\[1pt] \hline & & & & & & & & & & &  \\ [-1.2em]
$SU(2)_L$ & {\bf 2} & {\bf 1} & {\bf 1} & {\bf 2} & {\bf 1} & {\bf 2} & {\bf 2} & {\bf 1} & {\bf 1} & {\bf 1} & {\bf 1} \\ \hline & & & & & & & & & & &  \\ [-1em]& & & & & & & & & & & \\ [-1.5em]
$U(1)_Y$ & $\frac{1}{6}$ & $\frac{2}{3}$ & $-\frac{1}{3}$ & $-\frac{1}{2}$ & $-1$ & $-\frac{1}{2}$ & $-\frac{1}{2}$ & 0 & 0 & 0 & 0  \\[2pt]  \hline & & & & & & & & & & & \\ [-1em]
$U(1)'$ & $0$ & $0$ & $0$ & $0$ & $0$ & $1$ & $-1$ & $-1$ & $1$ & $\frac12$ & 1 \\ [0.2em] \hline 
\end{tabular}
\caption{\label{tab:Ib}Irreducible representations of the fields of the model under the electroweak $SU(2)_L \times U(1)_Y \times U(1)'$ gauge symmetry. The fields ${q_L}_{\alpha}, {\ell_L}_{\alpha}$ are left-handed SM doublets while ${u_R}_\beta,{d_R}_\beta,{e_R}_\beta$ are right-handed SM singlets where $\alpha, \beta$ label the three families of quarks and leptons. The fields $N_{\mathrm{R}{1,2}}$ are the two right-handed neutrinos.}
\end{table}

Here, we introduce the $U(1)'$ extension of minimal type Ib seesaw model with a Majorana fermion singlet. The charges of the fields in the model are summarised in Tab.\ref{tab:Ib}. The $U'(1)$ gauge symmetry, rather than any discrete $Z_3$ or $Z_4$ symmetries \cite{Chianese:2021toe,Fu:2021fyk}, is responsible for making the two Higgs doublets distinguishable and ensuring the type Ib seesaw structure. However, the $U(1)'$ symmetry does not completely take over the function of the discrete symmetries. In fact, the Yukawa interaction between charged fermions and Higgs doublets is forbidden by the $U(1)'$ symmetry. To preserve the fermion mass, a new scalar singlet $\phi$, which is also referred to as the ``Yukon'' \cite{King:2020mau}, is introduced with which dimension-5 interaction is allowed in the form of $\overline{q_L}_\alpha \Phi_2 u_{R\beta} \phi$, $\overline{q_L}_\alpha {\tilde \Phi}_1 d_{R\beta} \phi$ and $\overline{\ell}_\alpha {\tilde \Phi}_1 e_{R\beta} \phi$.\footnote{Dimension-5 operators $\overline{q_L}_\alpha \Phi_1 u_{R\beta} \phi^*$, $\overline{q_L}_\alpha {\tilde \Phi}_2 d_{R\beta} \phi^*$ and $\overline{\ell}_\alpha {\tilde \Phi}_2 e_{R\beta} \phi^*$ are also allowed by the $U(1)'$ gauge symmetry. However, to avoid flavour-changing neutral currents, we require those interactions to be forbidden. We will ignore these operators in the analysis of the effective theory, since they will be forbidden when we consider the more complete theory later.
} After $\phi$ gains a VEV $\langle\phi\rangle=v_\phi/\sqrt{2}$, the Yukawa interactions generating fermion mass after spontaneous symmetry breaking (SSB) of Higgs doublets are
\begin{eqnarray}
\mathcal{L}_{\rm 2HDM} & \supset & - Y^u_{\alpha \beta} \overline{q_L}_\alpha \Phi_2 u_{\mathrm{R}\beta}
- Y^d_{\alpha \beta} \overline{q_L}_\alpha {\tilde \Phi}_1 d_{\mathrm{R}\beta}- Y^e_{\alpha \beta} \overline{\ell}_\alpha {\tilde \Phi}_1 e_{\mathrm{R}\beta} + {\rm h.c.}\,,
\label{eq:Yuk} 
\end{eqnarray}
which is referred to as the type II two Higgs doublet model (2HDM) \cite{Branco:2011iw}. However, the $SU(2)\times U(1)\times U(1)'$ symmetry does not help to keep the type II 2HDM structure, which may lead to unexpected flavour changing process. As will be discussed in Sec.\ref{sec:four}, the problem can be solved by considering a fourth generation of vector-like fermions, which is also motivated by constructing a renormalisable model. After realising the type Ib seesaw model effectively, the type II 2HDM structure appears automatically.

Under the $U(1)'$ symmetry, the Yukawa interactions allowed in the type Ib seesaw sector take the form
\begin{eqnarray}
\mathcal{L}_{\rm seesawIb} & = & - Y_{1\alpha} \overline{\ell_L}_\alpha {\Phi}_1 N_{\mathrm{R}1} 
- Y_{2\alpha} \overline{\ell_L}_\alpha {\Phi}_2 N_{\mathrm{R}2}
- M\overline{N^c_{\mathrm{R}1}} N_{\mathrm{R}2} + {\rm h.c.}\,,
\label{eq:lg_Ib}
\end{eqnarray}
The two ``right-handed'' Weyl neutrinos can actually form a four component Dirac spinor $\N = \left(N^c_{R1},N_{R2} \right)$ with a Dirac mass $M$. The $U(1)'$ Dirac spinor $\N$ can be easily read as 1 from Tab.\ref{tab:Ib}. Notice that any Majorana mass terms of the RH neutrinos break the $U(1)'$ symmetry and therefore the classical type Ia seesaw is forbidden in this model. The type Ib seesaw Lagrangian can be rewritten in $\N$ as 
\begin{eqnarray}
\mathcal{L}_{\rm seesawIb} & = & - Y_{1\alpha}^* \overline{\ell^c_L}_\alpha \Phi_1^* \N_\mathrm{L} - Y_{2\alpha} \overline{\ell_L}_\alpha {\Phi}_2 \N_\mathrm{R} - M_N\overline{\N_\mathrm{L}} \N_\mathrm{R}+ {\rm h.c.}\,.
\label{eq:lagNS}
\end{eqnarray}

\begin{figure}[t!]
\begin{center}
\begin{fmffile}{FeynDiag/NM}
\fmfframe(10,10)(10,10){
\begin{fmfgraph*}(200,45)
\fmflabel{$\ell_\beta$}{o1}
\fmflabel{}{o2}
\fmflabel{}{i2}
\fmflabel{${\ell_L}_\alpha$}{i1}
\fmfv{}{v1}
\fmfv{decor.shape=cross,decor.size=8,label=$M_N$,label.angle=90}{v2}
\fmfv{}{v3}
\fmfv{label=$\Phi_1$,label.angle=90}{v4}
\fmfv{}{v5}
\fmfv{label=$\Phi_2$,label.angle=90}{v6}
\fmfleft{i1,i2}
\fmfright{o1,o2}
\fmf{fermion,tension=0.6}{i1,v1}
\fmf{fermion,label=$\N$,l.side=left}{v2,v1}
\fmf{fermion,label=$\N$,l.side=left}{v3,v2}
\fmf{fermion,tension=0.6}{o1,v3}
\fmf{phantom,tension=0.6}{i2,v4}
\fmf{phantom}{v5,v4}
\fmf{phantom}{v5,v6}
\fmf{phantom,tension=0.6}{o2,v6}
\fmf{dashes,tension=0}{v1,v4}
\fmf{phantom,tension=0}{v2,v5}
\fmf{dashes,tension=0}{v3,v6}
\end{fmfgraph*}}
\end{fmffile}
\label{fig:NMdirac}
\caption{Light neutrino mass generated by the type Ib seesaw mechanism}
\end{center}
\end{figure}
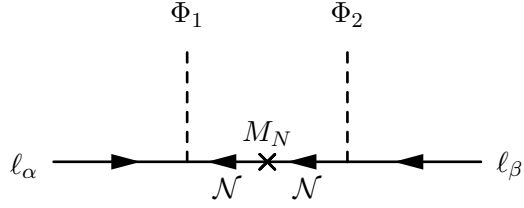

In Tab.\ref{tab:Ib}, the SM fermions are uncharged under $U(1)'$ to avoid chiral anomalies while the two Higgs doublets $\Phi_{1,2}$, the heavy neutrino $\N$, the dark fermion singlet $\chi_{L,R}$ and the Yukon $\phi$ are charged. The kinetic terms of those particles are 
\begin{eqnarray}
\mathcal{L}_{\rm U'(1)} =  \left(D'_\mu \Phi_1\right)^\dagger {D'}^\mu \Phi_1 + \left(D'_\mu \Phi_2\right)^\dagger  {D'}^\mu \Phi_2 &+& i\overline{\N} \slashed{D}' \N \nonumber\\&+& i\overline{\chi_L} \slashed{D}'\chi_L + i\overline{\chi_R} \slashed{D}'\chi_R + D'_\mu \overline{\phi} {D'}^\mu \phi 
\label{eq:lg_Ib}
\end{eqnarray}
where the covariant derivative under $SU(1)\times U(1)\times U(1)'$ symmetry is 
\begin{eqnarray}
&D'_\mu= \partial_\mu + i\frac12 g_2\,\boldsymbol{\sigma} \cdot \boldsymbol{W}_\mu + ig_1YB_\mu + ig'_1Y'B'_\mu\,.
\end{eqnarray}
In addition to the kinetic terms, the dark fermion can only couple to the Yukon through interaction
\begin{eqnarray}
y_\chi^L\overline{\phi}\,\overline{\chi_L^c} \chi_L + y_\chi^R\overline{\phi}\,\overline{\chi_R^c} \chi_R + h.c.\,.
\end{eqnarray}
After the $U(1)'$ is broken by the VEV of Yukon, the dark fermion $\chi$ gains a Majorana mass $m_{L,R}=\sqrt{2} y_\chi^{L,R} v_\phi$ and become stable due to its half-integer charge under $U(1)'$. In fact, the VEV of Yukon breaks the $U(1)'$ symmetry into a $Z_2$ symmetry under which $\chi$ is the only charged particle. In addition to the Majorana mass, the dark fermion can also have a Dirac mass $m_D$, which makes the mass matrix of $\chi_{L,R}$
\begin{eqnarray}
\begin{pmatrix} m_L & m_D \\ m_D & m_R \end{pmatrix}
\end{eqnarray}
Here, we consider two limits: hierarchical and degenerate dark fermion masses. The dark fermions appear to have a hierarchical mass spectrum when their Majorana masses are hierarchical and Dirac mass is negligible when compared to the heavier Majorana mass. Without loss of generality, we assume the left-handed dark singlet is heavier, i.e. $m_L\gg m_D,m_R$. Then $\chi_L$ is unstable and decays before the freeze-out of the stable dark matter candidate $\chi_R$. In this case, the existence of the left-handed dark fermion does not cause any significant effect in dark matter production. On the other hand, if the dark fermions share the same mass, either Dirac ($m_{L,R}=0$) or Majorana ($m_D=0$), the dark fermions are both stable and the predicted dark matter relic abundance is twice that in the case with hierarchical masses. Between these two limits, the dark fermion masses could be quasi-degenerate when $m_D\gg m_{L,R}\neq 0$ or $m_{L,R}\gg m_D\neq 0$ and the heavier dark fermion could have a long enough lifetime to play a role in dark matter production. For simplicity, we focus on the case where the dark fermion masses are hierarchical and only $\chi_R$ is considered during the freeze-out production of dark matter. In the rest of the paper, we adopt $\chi$ and $m_\chi$ for the dark matter candidate $\chi_R$ and its mass to simplify the notations.

Besides the dark fermion, the $U(1)'$ gauge boson $Z'$ also gains mass $M_{Z'}=g'_1v_\phi$ from the VEV of Yukon. However, since the Higgs doublets are also charged under the $U(1)'$ gauge symmetry, the mass of $Z'$ also receives contributions from $\langle\Phi_1\rangle$ and $\langle\Phi_2\rangle$ after the electroweak (EW) symmetry breaking, which leads to mixing between the massive gauge bosons.

\subsection{Gauge boson mixing}
After the SSB of the Higgs doublets, the mass matrix of $W_3$, $B$ and $B'$ reads
\begin{eqnarray}
\frac{v^2}{4}\begin{pmatrix} 
g_2^2 & - g_1g_2  & 2g_2g'_1\cos2\beta\\
- g_1g_2 & g_1^2 & -2g_1g'_1\cos2\beta\\
2g_2g'_1\cos2\beta & -2g_1g'_1\cos2\beta & 4{g'_1}^2 \left(1+ \frac{v_\phi^2}{v^2}\right)
\end{pmatrix}
\end{eqnarray}
where $\beta=\arctan\left(\langle\Phi_2\rangle/\langle\Phi_2\rangle\right)$. The mass of $B'$ also receives a contribution from the SSB of scalar singlet $\phi$ as $M_{Z'}=g'_1v_\phi$. While the photon remains massless, there is a mixing between the SM neutral gauge boson $Z$ and the gauge boson $Z'$ 
\begin{eqnarray}
Z\rightarrow Z \cos\theta - Z'\sin\theta\,, \quad
Z'\rightarrow Z' \cos\theta + Z\sin\theta
\end{eqnarray}
with the expression of the mixing angle given by 
\begin{eqnarray}
\tan2\theta = \frac{2\cos2\beta\, g'_1 v\, M_Z}{M_{Z'}^2-M_Z^2+{g'_1}^2 v^2} 
= \frac{4\cos2\beta\, g'_1 \sqrt{g_1^2+g_2^2} v^2}{4{g'_1}^2 (v^2+v_\phi^2)-(g_1^2+g_2^2) v^2}\,.
\end{eqnarray}
Assuming $M_{Z'}\gg M_Z$ and $v_\phi \gg v$, the mixing angle $\theta$ is approximately $\cos2\beta\, \sqrt{g_1^2+g_2^2}\, v^2 / g'_1 v_\phi^2$. The EW Precision Observables provides an upper bound on the mixing angle, which is $\theta \lesssim 10^{-3}$ \cite{DELPHI:1994ufk}. The upper bound can be converted into constraint on the parameters in the model as 
\begin{eqnarray}
g'_1 v_\phi^2 \gtrsim \left( 6.7\, \text{TeV} \right)^2\,.
\label{eq:mixing}
\end{eqnarray}
Due to the perturbativity limit of the $U(1)'$ gauge coupling $g'_1$, $v_\phi$ has to be larger than $3.6$ TeV, which is coincident with the assumption $v_\phi \gg v$ above.

\begin{figure}[t!]
\begin{center}
\begin{fmffile}{FeynDiag/DMpro1}
\fmfframe(20,20)(20,20){
\begin{fmfgraph*}(80,45)
\fmflabel{$\chi$}{o1}
\fmflabel{$\chi$}{o2}
\fmflabel{$\N$}{i2}
\fmflabel{$\overline{\N}$}{i1}
\fmfv{label=$g_1'$}{v1}
\fmfv{label=$\frac{g_1'}{2}$}{v2}
\fmfleft{i1,i2}
\fmfright{o1,o2}
\fmf{plain}{i2,v1}
\fmf{plain}{i1,v1}
\fmf{plain}{v2,o1}
\fmf{plain}{v2,o2}
\fmf{photon,label=$Z'$}{v1,v2}
\fmfdotn{v}{2}
\end{fmfgraph*}}
\end{fmffile}
\begin{fmffile}{FeynDiag/DMpro2}
\fmfframe(20,20)(20,20){
\begin{fmfgraph*}(80,45)
\fmflabel{$\chi$}{o1}
\fmflabel{$\chi$}{o2}
\fmflabel{$\overline{\Phi}_i$}{i1}
\fmflabel{$\Phi_i$}{i2}
\fmfv{label=$g_1'$}{v1}
\fmfv{label=$\frac{g_1'}{2}$}{v2}
\fmfleft{i1,i2}
\fmfright{o1,o2}
\fmf{dashes}{i2,v1}
\fmf{dashes}{i1,v1}
\fmf{plain}{v2,o1}
\fmf{plain}{v2,o2}
\fmf{photon,label=$Z'$}{v1,v2}
\fmfdotn{v}{2}
\end{fmfgraph*}}
\end{fmffile}
\end{center}
\caption{\label{fig:Feyn} The processes responsible for DM production}
\end{figure}
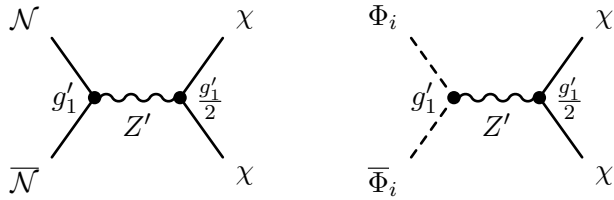

\subsection{Freeze-out production of dark matter }

In the early universe, the dark matter candidate $\chi$ can interact with the other particles through the $Z'$ mediated processes as shown in Fig.\ref{fig:Feyn}.\footnote{The $U(1)'$ gauge boson and the Yukon are assumed to be decoupled at the freeze-out temperature. The gauge boson $Z'$ decays fast because of the large gauge coupling. More specifically, in the region of parameter space $m_{Z'} \sim m_\chi$, the decay rate of $Z'$ roughly reads ${g'_1}^2M_{Z'}/8\pi$. The ratio of the decay rate and Hubble rate reads
\begin{eqnarray}
\Gamma_{Z'}/\Hub \simeq 0.002\,{g'_1}^2\frac{M_{Z'}M_{\rm P}} {T_f^2}\nonumber \,,
\end{eqnarray}
where $M_{\rm P}$ is the Planck mass. Due to the large Planck mass, the gauge boson $Z'$ is likely to decouple at temperature higher than $M_{Z'}$ when $Z'$ is TeV scale. On the other hand, the freeze-out temperature is typically 20 times smaller than $m_\chi$. Therefore $Z'$ decouples before the freeze-out takes place. And the Yukon is also decoupled for similar reasons.} Since the vertices all involve gauge couplings, we do not expect the interaction to be feeble and therefore consider the freeze-out production of the dark fermion $\chi$. The Boltzmann equation of $\chi$ is \cite{Kolb:1990vq}
\begin{eqnarray}
\frac{d Y_\chi}{d X} & = & 
- \frac{ X \mathfrak{s} }{\Hub(m)} \left<\sigma\, v\right>_{\chi\chi} \left(Y_\chi^2-{Y_\chi^{\rm eq}}^2\right)\,, \label{eq:BE}
\end{eqnarray}
where $X\equiv m_\chi/T$. The quantities $\mathcal{H}$ and $\mathfrak{s}$ are the Hubble parameter and the entropy density of the thermal bath. The comoving density $Y_\chi$ is defined as the ratio of number density and entropy density $n_\chi/\mathfrak{s}$. The superscript ``eq'' represents the value of the quantity in thermal equilibrium. Define $X_f$ as the ratio of DM mass $m_\chi$ and the freeze-out temperature $T_f$. To provide an analytical view of the solution to the Boltzmann equation, we consider both the observational constraint and the general feature of freeze-out. On the one hand, by requiring the correct DM relic abundance, i.e. $Y_{\rm DM}^{\rm obs} = Y_{\chi}^{\rm eq}(X_f)$, the observed DM comoving density can be computed from 
\begin{equation}
\Omega_{\rm DM}^\mathrm{obs}h^2 =  \frac{\mathfrak{s}_0 \, m_\chi  \,Y_{\rm DM}^{\rm obs}}{\rho_{\rm crit}/h^2}  \,,
\label{eq:omegaPREA}
\end{equation}
where $\mathfrak{s}_0=2891.2\,{\rm cm^3}$ and $\rho_{\rm crit}/h^2 = 1.054 \times10^{-5} {\rm GeV \, cm^{-3}}$ are the current entropy density and the critical density, respectively \cite{ParticleDataGroup:2018ovx}. The DM relic abundance is measured by the Planck Collaboration at 68\% C.L. \cite{Aghanim:2018eyx}
\begin{equation}
\Omega_{\rm DM}^\mathrm{obs}h^2 = 0.120 \pm 0.001\,.
\label{eq:omegaOBS}
\end{equation}
At the time of freeze-out, the comoving density of $\chi$ in thermal equilibrium takes the expression
\begin{equation}
Y^{\rm eq}_\chi \equiv \frac{n^{\rm eq}_\chi}{\mathfrak{s}} = \frac{45 g_\chi }{4\pi^4 g^\mathfrak{s}_*} X_f^2 K_2\left(X_f\right)\,,
\label{eq:Yield_chi}
\end{equation}
where $g_\chi$ is the degree of freedom in $\chi$ and $g^\mathfrak{s}_*$ is the degree of freedom of the relativistic species in the thermal bath. The function $K_2$ is the modified Bessel function of the second kind with order 2.
Simple calculation shows that $X_f$ has to satisfy
\begin{eqnarray}
X_f \equiv \frac{m_\chi}{T_f} \simeq 27.4 + 1.07\ln \frac{m_\chi}{1 \text{TeV}}\,.\label{eq:Xf}
\end{eqnarray}
On the other hand, as a general feature of thermal production, a particle decouples from the thermal bath when the rate of its interaction with particles in the thermal bath drops below the Hubble constant. This freeze-out criterion indicates that $\Gamma_\chi(X_f)\simeq \Hub(X_f)$. On the left side, the expression of the interaction rate reads $\Gamma_\chi=\left<\sigma\, v\right>_{\chi\chi} n^{\rm eq}_\chi$ and the number density of $\chi$ can be computed from Eq.\eqref{eq:omegaPREA} and  Eq.\eqref{eq:Yield_chi} as
\begin{equation}
n^{\rm eq}_{\chi} = Y^{\rm eq}_\chi \mathfrak{s} =  \frac{\Omega_{\rm DM}^\mathrm{obs} \rho_{\rm crit}}{\mathfrak{s}_0} \frac{2\pi^2}{45} g^\mathfrak{s}_*\left(T_f\right) X_f T_f^2 \,.
\label{eq:neq}
\end{equation}
 On the right side, the temperature dependent expression of the Hubble constant is 
\begin{equation}
\mathcal{H} (T)= \sqrt{\frac{4\pi^3{g_*}\left(T_f\right)}{45}}\frac{T_f^2}{M_{\rm P}} \,,
\label{eq:Hubbel_constant}
\end{equation}
where $g_*$ is the degrees of freedom of the relativistic species in the thermal bath and $M_{\rm P}$ is the Planck mass. As a consequence, the freeze-out criterion leads to
\begin{eqnarray}
\left<\sigma\, v\right>_{\chi\chi}X_f^{-1}= 6.63 \times 10^{-5}\, \text{TeV}^{-2}\,.\label{fo_crit}
\end{eqnarray} 
The thermally averaged cross section $\left<\sigma\, v\right>_{\chi\chi}$ can be computed using a general expression \cite{Edsjo:1997bg}
\begin{equation}
\left<\sigma_{ij\rightarrow kl}\, v_{ij}\right> = \frac{1}{n^{\rm eq}_{i}\, n^{\rm eq}_{j}}\frac{g_i \, g_j}{S_{kl}}\frac{T}{512\pi^6} \int_{\left(m_i+m_j\right)^2}^\infty ds \, \frac{p_{ij} \, p_{kl} \,K_1\left(\sqrt{s}/T\right)}{\sqrt{s}} \int  \overline{\left|\mathcal{M}\right|^2}_{ij\rightarrow kl} \, d\Omega\,.
\label{avgsection}
\end{equation}
where $K_1$ is the order-1 modified Bessel function of the second kind. The quantities $s$, $S_{kl}$ and $p_{ij}$ ($p_{kl}$) are the square of the centre-of-mass energy, the symmetry factor and the initial (final) centre-of-mass momentum, respectively. The scattering amplitudes of the processes in Fig.\ref{fig:Feyn} read
\begin{eqnarray}
\int \overline{\left|\mathcal{M}\right|^2}_{\chi\chi \rightarrow \N\overline{\N}}\, d\Omega &=& {g'_1}^4\frac{2\pi}{3} \frac{(s+2M^2)(s-m_\chi^2)}{(s-M_{Z'}^2)^2 + M_{Z'}^2 \Gamma_{Z'}^2}\,,\\
\int \overline{\left|\mathcal{M}\right|^2}_{\chi\chi \rightarrow \Phi_{i}\overline{\Phi}_{i}}\, d\Omega &=& {g'_1}^4 \frac{\pi}{6} \frac{(s-4M_{\Phi_i}^2)(s-m_\chi^2)}{(s-M_{Z'}^2)^2 + M_{Z'}^2 \Gamma_{Z'}^2}\,.
\end{eqnarray}
Prompted by one of the motivations of the type Ib seesaw mechanism, the Dirac neutrino is assumed to be around 1-100 GeV scale, where a testable dark matter model compatible with leptogenesis can be realised \cite{Chianese:2021toe,Fu:2021fyk}. In this research, we focus on dark matter candidates above TeV scale, which means $m_\chi \gg M, M_{\Phi_i}$\footnote{The case where the RH neutrinos are heavier than dark matter candidates has been studied generally in \cite{Blennow:2019fhy}.} and the total averaged cross section reads
\begin{eqnarray}
\left<\sigma\, v\right>_{\chi\chi} &=& \frac{g_\chi^2}{(n^{\rm eq}_\chi)^2}\frac{T}{2048\pi^5} \, {g'_1}^4 \int_{4m_\chi^2}^\infty ds \,\sqrt{s-4m_\chi^2}\,K_1\left(\sqrt{s}/T\right)\frac{s(s-m_\chi^2)}{(s-M_{Z'}^2)^2  + M_{Z'}^2 \Gamma_{Z'}^2}\,.\quad\quad
\label{avgsection_chi}
\end{eqnarray}
The decay rate of $Z'$ depends on the mass of $Z'$ and gauge coupling $g'_1$ as
\begin{eqnarray}
\Gamma_{Z'}={g'_1}^2\begin{dcases} 
\frac{1}{8\pi}M_{Z'} & M_{Z'}<2m_\chi\,, \\
\frac{1}{8\pi}M_{Z'} +\frac{1}{96\pi}\frac{M_{Z'}^2-m_\chi^2}{M_{Z'}^2}\sqrt{M_{Z'}^2-4m_\chi^2} & M_{Z'}>2m_\chi\,. \\
\end{dcases}
\end{eqnarray}
There are two different scenarios distinguished by whether the gauge boson $Z'$ can decay into two dark fermions or not. If the decay $Z'\rightarrow\chi\chi$ is allowed, i.e. $M_{Z'}>2m_\chi$, a resonance can be observed in Eq.\eqref{avgsection_chi}.

To obtain analytical results, we consider two different limits in the masses of the dark fermion and the $U(1)'$ gauge boson: $M_{Z'}\gg m_\chi$ and $M_{Z'}\ll m_\chi$. In the case that $M_{Z'}\gg m_\chi$, the decay process of $Z'$ into dark fermions is kinetically allowed and the total averaged cross section can be simplified into
\begin{eqnarray}
\left<\sigma\, v\right>_{\chi\chi} &=&  \frac{3{g'_1}^4}{64\pi M_{Z'}^4}\left(1+\frac{169}{9216\pi^2}{g'_1}^4 \right)^{-1}\left[m_\chi^2 \frac{K_1^2}{K_2^2} +4m_\chi T\frac{K_1}{K_2}+ \left(m_\chi^2 + 8T^2 \right)\right]\,.
\end{eqnarray}
Around the freeze-out temperature, Eq.\eqref{eq:Xf} implies $T\ll m_\chi$ and therefore
\begin{eqnarray}
\left<\sigma\, v\right>_{\chi\chi} &\simeq& 
\frac{3{g'_1}^4m_\chi^2}{32\pi M_{Z'}^4}\left(1+\frac{169}{9216\pi^2}{g'_1}^4 \right)^{-1} = 
\frac{3m_\chi^2}{32\pi v_\phi^4} \left( 1+\frac{169}{1024\pi^2}{g'_1}^4 \right)^{-1}\,.
\end{eqnarray}
By applying the freeze-out criterion in Eq.\eqref{fo_crit}, it can be derived that 
\begin{eqnarray}
v_\phi^2 \simeq \frac{m_\chi}{1\, \text{TeV}}\frac{(4.6\, \text{TeV})^2}{\sqrt{27.4 + 1.07\ln \frac{m_\chi}{1 \text{TeV}}}} \left( 1+\frac{169}{9216\pi^2}{g'_1}^4 \right)^{-1/2}\,
\end{eqnarray}
However, the gauge coupling $g'_1$ has to be below its perturbativity limit and therefore the assumption $M_{Z'}=g'_1v_\phi \gg m_\chi$ requires
\begin{eqnarray}
v_\phi \ll 7.1\, \text{TeV}\,.
\end{eqnarray}
Remember that there is also a lower bound on the value of $v_\phi$ from the gauge boson mixing strength, which is about 3.6 TeV. Thus the limit $M_{Z'}\gg m_\chi$ is not eligible. 

On the other hand, in the case of $M_{Z'} \ll m_\chi$, the thermal average cross section can be computed as 
\begin{eqnarray}
\left<\sigma\, v\right>_{\chi\chi} &=&  \frac{3{g'_1}^4}{512\pi m_\chi^2}
\end{eqnarray}
when the temperature is much lower than the dark fermion mass ($T\ll m_\chi$). Then the freeze-out criterion Eq.\eqref{fo_crit} leads to 
\begin{eqnarray}
{g'_1}^4 m_\chi^{-2} X_f^{-1} &\simeq& 3.55 \times 10^{-2}\, \text{TeV}^{-2}\,.
\end{eqnarray}
With the approximated expression of $X_f$ in Eq.\eqref{eq:Xf}, the result can be further simplified into the relation between the $U(1)'$ gauge coupling and dark fermion mass
\begin{eqnarray}
{g'_1}^2&\simeq& 0.19\frac{m_\chi}{1\,\text{TeV}}\sqrt{27.4 + 1.07\ln \frac{m_\chi}{1 \text{TeV}}}\,.
\end{eqnarray}
Again, by considering the perturbative limit of $g'_1$, the dark matter mass has to satisfy $m_\chi < 12.2\, \text{TeV}$. 
However, the constraint on the gauge boson mixing requires $v_\phi$ to be at least $3.6$ TeV. Therefore the gauge boson mass $M_{Z'}$ has a lower bound 
\begin{eqnarray}
M_{Z'}\gtrsim{g'_1} \times 3.6\,\text{TeV} &\simeq& 1.5\,\text{TeV}\sqrt{\frac{m_\chi}{1\,\text{TeV}}\sqrt{27.4+ 1.07\ln \frac{m_\chi}{1 \text{TeV}}}}\,.
\end{eqnarray}  
From this lower bound, it can be easily derived that the gauge boson $Z'$ cannot be lighter than the dark fermion $\chi$ if the dark fermion mass $m_\chi$ is less than $13.1$ TeV. Moreover, the ratio $M_{Z'}/m_\chi$ cannot be less than 1 for $m_\chi < 12.2\, \text{TeV}$, which is required by the perturbativity of $g'_1$. Therefore, again, the assumption of this scenario is broken and the scenario is forbidden by the constraint from the gauge boson mixing.

\begin{figure}[t!]
\begin{center}
\subfigure[]{\includegraphics[width=0.49\textwidth]{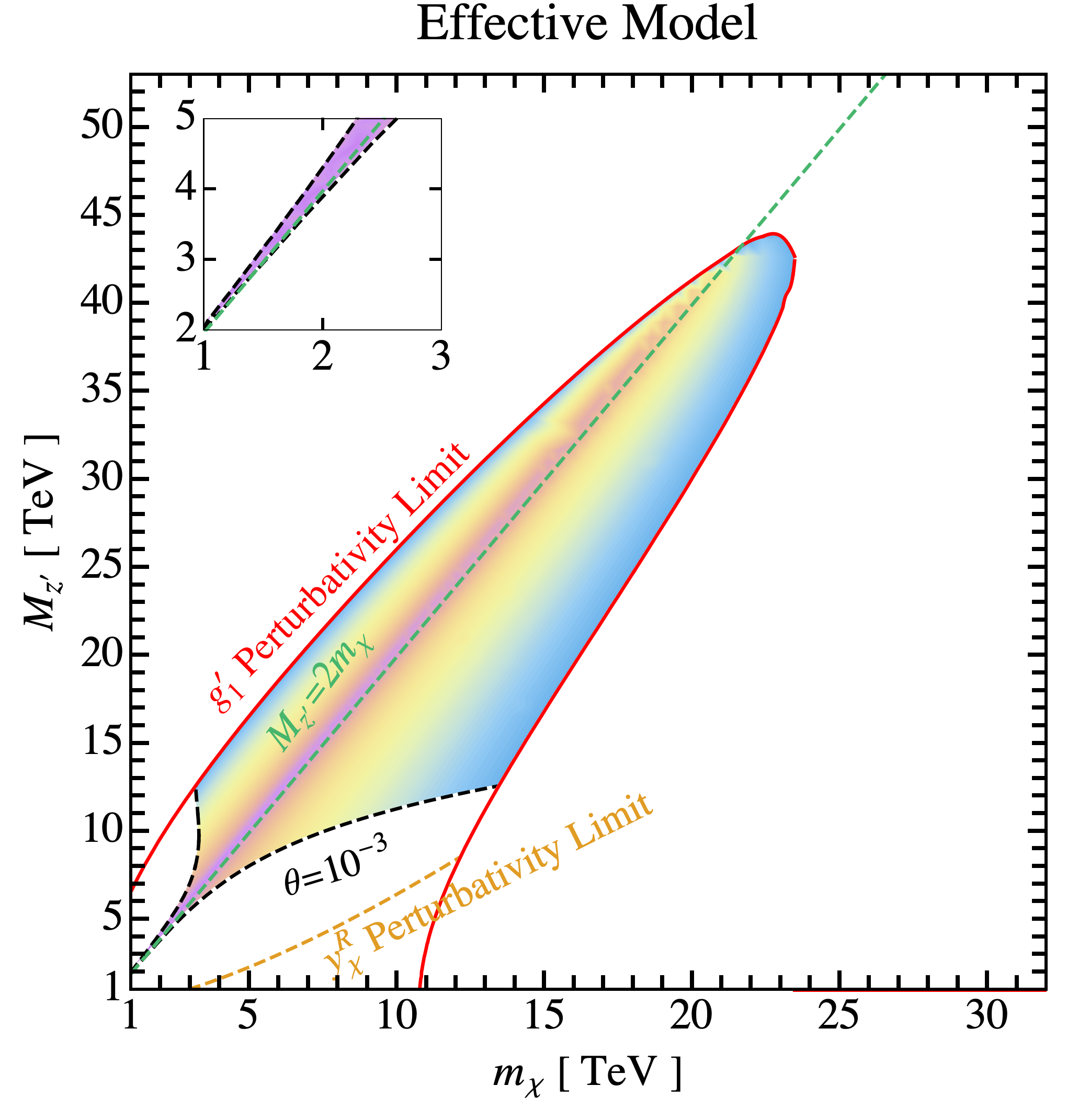}\label{fig:ra_eff}}
\subfigure[]{\includegraphics[width=0.49\textwidth]{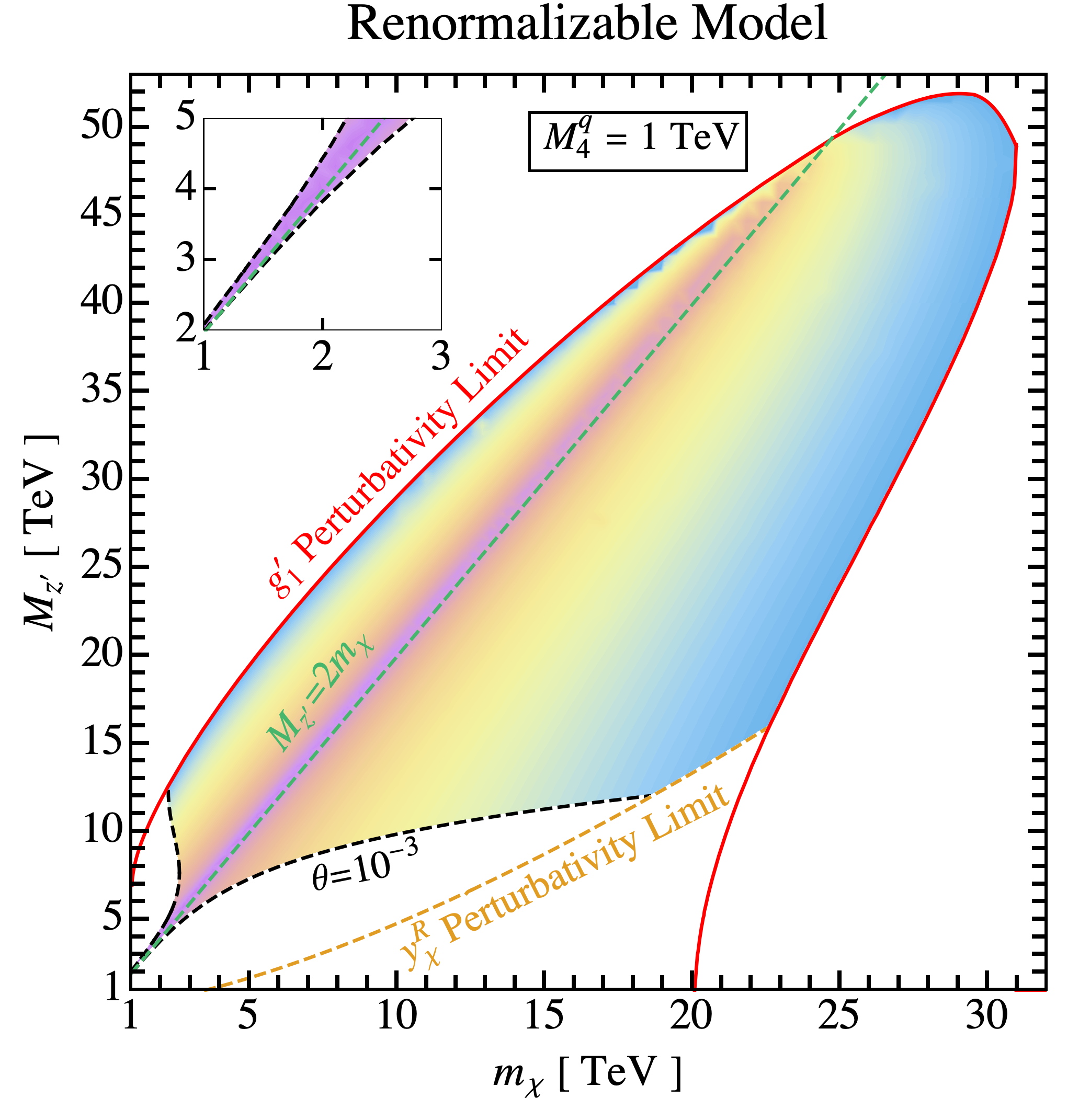}\label{fig:ra_4th}}
\includegraphics[width=0.49\textwidth]{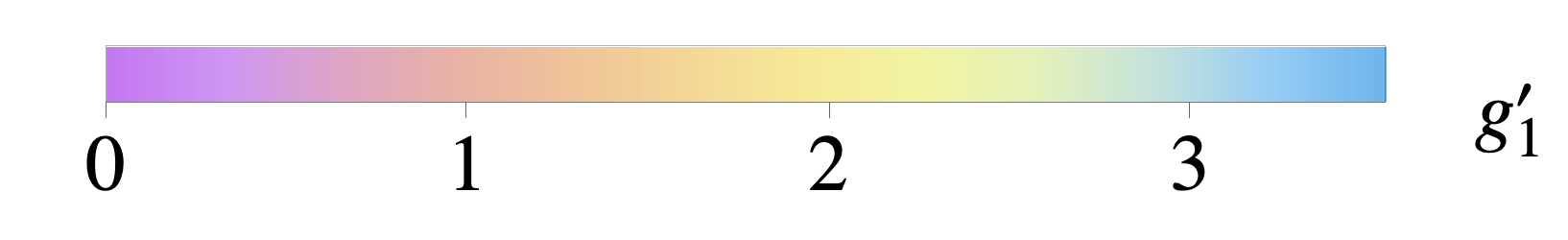}
\caption{ Allowed values of $U(1)'$ gauge coupling for different masses of dark fermion and $U(1)'$ gauge boson in the effective model (left panel) and renormalisable model. }
\end{center}
\end{figure}

Since the scenarios where $M_{Z'} \gg m_\chi$ and $M_{Z'} \ll m_\chi$ have been proved to be illegal, the only remaining case is $M_{Z'} \sim m_\chi$. For such a scenario, an analytical calculation is hard to be performed so some numerical results are shown in Fig.\ref{fig:ra_eff}. Similar to the analytical derivation, the numerical solution of the gauge coupling is obtained from the Boltzmann equation by requiring the correct dark matter relic abundance for each pair of dark fermion and gauge boson masses $\left(m_\chi,\,M_{Z'}\right)$. By neglecting the GeV scale neutrino mass, the remaining free parameters affecting dark matter production are the $U(1)'$ gauge coupling $g'_1$, the dark fermion mass $m_\chi$ and the $U(1)'$ gauge boson mass $M_{Z'}$. The allowed parameter space is coloured by the required coupling constant $g'_1$ in the plot, while the excluded region is left unfilled. The red solid and yellow dashed lines in the figure show the perturbativity limits of the gauge coupling $g'_1$ and the Yukawa coupling $y_\chi^R$, respectively. The constraint from gauge boson mixing is presented as the black dashed line. Besides, the threshold of two different scenarios of the $Z'$ decay is also marked out by a green dashed line. As has been proved analytically, correct relic abundance can only be produced when the masses of the dark fermion $\chi$ and the $U(1)'$ gauge boson $Z'$ are similar. In the allowed region in the parameter space, the VEV of the scalar singlet, which is given by $v_\phi=M_{Z'}/g'_1$, is always far larger than the freeze-out temperature that is typically 20 times below the dark fermion mass $m_\chi$. As a result, no significant thermal effect contributes to the total cross section. 

From Fig.\ref{fig:ra_eff}, it can be easily figured out that the masses of $\chi$ and $Z'$ cannot exceed 24 TeV and 44 TeV respectively, while the parameter space for $m_\chi<1$ TeV and $m_{Z'}<2$ TeV is very unfavored by the massive gauge boson mixing angle. The required gauge coupling is relatively small along the threshold line for different scenarios of the $Z'$ decay with a minimum value around 0.04. Above the threshold line, the resonance is statistically suppressed during the freeze-out, while no resonance appears below the threshold line. As the $U(1)'$ gauge boson mass decreases, the coupling $g'_1$ needs to be smaller as required by the gauge boson mixing in Eq.\eqref{eq:mixing}. When both the dark fermion $\chi$ and the $U(1)'$ gauge boson are a few TeV, the required coupling $g'_1$ is so small that the correct dark matter density can only be produced through the resonance in the propagator. As a result, the gauge boson mass $M_{Z'}$ is nearly twice the mass of the dark fermion $\chi$, which is also shown in the zoomed-in subfigure on the top left corner. The perturbativity limit of the Yukawa coupling $y_\chi^R$,\footnote{Notice that the perturbativity limit of $y_\chi^R$ cannot be actually achieved since we are in a regime where $m_L \gg m_R$, which means the perturbativity limit of $y_\chi^L$ is a stronger constraint on the parameter space. However, as the ratio between the Majorana masses of the dark fermions is randomly large, we only show the constraint from the perturbativity limit of $y_\chi^R$ as a guidance.} although is shown in this figure, is always weaker than the constraint from massive gauge boson mixing and does not help to constrain the parameter space.

\subsection{Dark matter detections}

Since the neutrino EW eigenstates are not charged under the $U(1)'$ gauge symmetry, the DM annihilation into neutrinos has to be approached either by the mixing of neutrino mass eigenstates or by the mixing between massive neutral gauge bosons. The mixing angle between the neutrinos is given by $m_{\rm D}M_N^{-1}$, where $(m_{\rm D})_{i\alpha}=Y_{i\alpha}v_i/\sqrt2$ and $M_N$ is the mass matrix of the Dirac neutrino, while the mixing angle between the massive gauge bosons is given in Eq.\eqref{eq:mixing} and below. For sub-TeV scale Dirac neutrinos, the experimental upper bound on the largest Yukawa coupling is typically around $0.01$ \cite{Chianese:2021toe}, therefore the active-sterile neutrino mixing can play an important role in indirect detection. At zero temperature limit, the cross section of DM annihilation is given by 
\begin{eqnarray}
\sigma_{\rm ann}&=&\frac{{g'_1}^4}{8\pi}\frac{m_\chi^2}{(4m_\chi^2-M_{Z'}^2)^2+ M_{Z'}^2 \Gamma_{Z'}^2}\left[\sum_{\alpha}\left(\sum_{i}\frac{Y_{i\alpha}v_i}{M} \right)^2 + \left(\frac{0.2 \,\text{TeV} }{M_{Z'}}\right)^4\right]\,.
\end{eqnarray}
Around the scale of a few TeV, the dark fermion and $U(1)'$ gauge boson masses roughly follow the relation $M_{Z'}=2m_\chi$, and the velocity averaged DM annihilation cross section can be expressed as 
\begin{eqnarray}
\langle\sigma_{\rm ann}v\rangle&=&1.2 \times10^{-25} \,\text{cm}^3/\text{s}\, \left(\frac{1 \,\text{TeV}}{M_{Z'}}\right)^2 \left[\sum_{\alpha}\left(\sum_{i}\frac{Y_{i\alpha}v_i}{M} \right)^2 + \left(\frac{0.2 \,\text{TeV} }{M_{Z'}}\right)^4\right]\,,
\end{eqnarray}
where the RMS velocity of the standard DM halo and the solar rotation speed are considered as in \cite{Frankiewicz:2015zma}. Inside the bracket, the first term is the contribution from active-sterile neutrino mixing, which is roughly constrained to be less than $10^{-5}$ by collider data \cite{Chianese:2021toe}. The second term in the bracket is the contribution from massive gauge boson mixing, which is constrained to be less than $10^{-4}$ as $Z'$ is required to be heavier than 2 TeV to produce the correct relic abundance as shown in figure Fig.\ref{fig:ra_eff}. Therefore the minimum required sensitivity in the velocity averaged DM annihilation cross section is around $10^{-29} \,\text{cm}^3/\text{s}$, which is lower than the expected sensitivity of Hyper-K\cite{Frankiewicz:2015zma}.

As in the case of DM annihilation, dark matter can interact with the nucleons through the mixing of massive gauge bosons.\footnote{In the renormalisable model discussed in the next section, the dark matter can also interact with the nucleons through quark mixing with the fourth family quarks. However, such a mixing only appears for the $d$ quark in nucleons as the $u$ quark has zero couplings to the fourth family quarks, and the strength is suppressed by the mass of $d_4$. } The cross section of DM-nucleon scattering has both a spin-independent (SI) component and a spin-dependent (SD) component. In this model, their expressions read \cite{Boveia:2016mrp}
\begin{eqnarray}
\sigma_{\rm SI}&=&5.3\times10^{-49}\,\text{cm}^2 {g'_1}^4\left(\frac{1\,\text{TeV}}{M_{Z'}}\right)^8\, \quad \text{and}\quad
\sigma_{\rm SD}=1.9\times10^{-47}\,\text{cm}^2 {g'_1}^4\left(\frac{1\,\text{TeV}}{M_{Z'}}\right)^8\,.
\end{eqnarray}
Due to Eq.\eqref{eq:mixing}, the SI component and SD component have to be smaller than $1.3\times10^{-55}$ and $4.6\times10^{-54}$ respectively, which lie below the current sensitivity of direct detection experiments \cite{XENON:2018voc,XENON:2019rxp,PICO:2019vsc,XENON:2020kmp}.

\section{Renormalisable model with fourth family vector-like fermions \label{sec:four}}

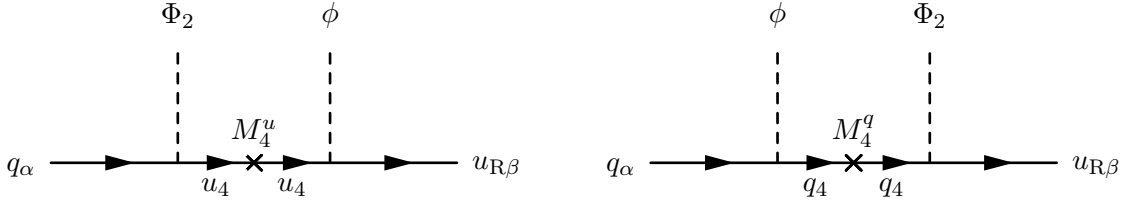
\begin{figure}[t!]
\begin{center}
\begin{fmffile}{FeynDiag/u_quark_Yuk_1}
\fmfframe(0,10)(10,10){
\begin{fmfgraph*}(190,45)
\fmflabel{$u_{\mathrm{R}\beta}$}{o1}
\fmflabel{}{o2}
\fmflabel{}{i2}
\fmflabel{$q_\alpha$}{i1}
\fmfv{}{v1}
\fmfv{decor.shape=cross,decor.size=8,label=$M^u_4$,label.angle=90}{v2}
\fmfv{}{v3}
\fmfv{label=$\Phi_2$,label.angle=90}{v4}
\fmfv{}{v5}
\fmfv{label=$\phi$,label.angle=90}{v6}
\fmfleft{i1,i2}
\fmfright{o1,o2}
\fmf{fermion,tension=0.6}{i1,v1}
\fmf{fermion,label=$u_4$,l.side=right}{v1,v2}
\fmf{fermion,label=$u_4$,l.side=right}{v2,v3}
\fmf{fermion,tension=0.6}{v3,o1}
\fmf{phantom,tension=0.6}{i2,v4}
\fmf{phantom}{v5,v4}
\fmf{phantom}{v5,v6}
\fmf{phantom,tension=0.6}{o2,v6}
\fmf{dashes,tension=0}{v1,v4}
\fmf{phantom,tension=0}{v2,v5}
\fmf{dashes,tension=0}{v3,v6}
\end{fmfgraph*}}
\end{fmffile}
\begin{fmffile}{FeynDiag/u_quark_Yuk_2}
\fmfframe(10,0)(10,10){
\begin{fmfgraph*}(190,45)
\fmflabel{$u_{\mathrm{R}\beta}$}{o1}
\fmflabel{}{o2}
\fmflabel{}{i2}
\fmflabel{$q_\alpha$}{i1}
\fmfv{}{v1}
\fmfv{decor.shape=cross,decor.size=8,label=$M^q_4$,label.angle=90}{v2}
\fmfv{}{v3}
\fmfv{label=$\phi$,label.angle=90}{v4}
\fmfv{}{v5}
\fmfv{label=$\Phi_2$,label.angle=90}{v6}
\fmfleft{i1,i2}
\fmfright{o1,o2}
\fmf{fermion,tension=0.6}{i1,v1}
\fmf{fermion,label=$q_4$,l.side=right}{v1,v2}
\fmf{fermion,label=$q_4$,l.side=right}{v2,v3}
\fmf{fermion,tension=0.6}{v3,o1}
\fmf{phantom,tension=0.6}{i2,v4}
\fmf{phantom}{v5,v4}
\fmf{phantom}{v5,v6}
\fmf{phantom,tension=0.6}{o2,v6}
\fmf{dashes,tension=0}{v1,v4}
\fmf{phantom,tension=0}{v2,v5}
\fmf{dashes,tension=0}{v3,v6}
\end{fmfgraph*}}
\end{fmffile}
\caption{\label{fig:u-quark} Effective interaction between quark doublets and up-type quarks}
\end{center}
\end{figure}

As mentioned before, the minimal model in Tab.\ref{tab:Ib} does not allow renormalisable interaction as the origin of charged fermion masses. Besides, the type II 2HDM structure is not ensured by any symmetry in the dimension-5 operators. Therefore, for the completeness of the theory, it is urgent to construct a renormalisable theory which preserves the structure of type II 2HDM in a natural way. This can be achieved by introducing a fourth family of vector-like fermions as shown in Tab.\ref{tab:Ib_ex}.
\begin{table}[t!]
\centering
\begin{tabular}{|c|c|c|c|c|c|c|c|c|c|c|c|c|c|c|c|c|c|}
\hline 
& ${q_L}_\alpha$ & ${u_R}_\beta$ & ${d_R}_\beta$ & ${\ell_L}_\alpha$ & ${e_R}_\beta$ & ${q}_4$ & $u_4$ & $d_4$ & $\ell_4$ & $e_4$ & $\Phi_{1}$ & $\Phi_{2}$ & $\N$ & $\chi_{L,R}$ & $\phi$\\[1pt] \hline & & & & & & & & & & & & & & &  \\ [-1.2em]
$SU(2)_L$ & {\bf 2} & {\bf 1} & {\bf 1} & {\bf 2} & {\bf 1} & {\bf 2} & {\bf 1} & {\bf 1} & {\bf 2} & {\bf 1} & {\bf 2} & {\bf 2} & {\bf 1} & {\bf 1} & {\bf 1} \\ \hline & & & & & & & & & & & & & & &  \\ [-1em]& & & & & & & & & & & & & & &  \\ [-1.5em]
$U(1)_Y$ & $\frac{1}{6}$ & $\frac{2}{3}$ & $-\frac{1}{3}$ & $-\frac{1}{2}$ & $-1$ & $\frac{1}{6}$ & $\frac{2}{3}$ & $-\frac{1}{3}$ & $-\frac{1}{2}$ & $-1$ & $-\frac{1}{2}$ & $-\frac{1}{2}$ & 0 & 0 & 0  \\[2pt]  \hline & & & & & & & & & & & & & & &  \\ [-1em]
$U(1)'$ & $0$ & $0$ & $0$ & $0$ & $0$ & $-1$ & $1$ & $1$ & $-1$ & $1$ & $1$ & $-1$ & $1$ & $\frac12$ & 1 \\ [0.2em] \hline 
\end{tabular}
\caption{\label{tab:Ib_ex}Irreducible representations of the fields of the model under the electroweak $SU(2)_L \times U(1)_Y \times U(1)'$ gauge symmetry. The fields ${q_L}_{\alpha}, {\ell_L}_{\alpha}$ are left-handed SM doublets while ${u_R}_\beta,{d_R}_\beta,{e_R}_\beta$ are right-handed SM singlets where $\alpha, \beta$ label the three families of quarks and leptons. The two right-handed neutrino fields $N_{\mathrm{R}{1,2}}$ are written as a Dirac pair $\N$.}
\end{table}
With new vector-like fermions, the allowed Yukawa interactions between charged fermions and scalars are 
\begin{eqnarray}
\mathcal{L}_{\rm Yuk} & \supset & 
- Y^{qu}_{\alpha 4} \overline{q_L}_\alpha \Phi_2 u_4 
- Y^{qd}_{\alpha 4} \overline{q_L}_\alpha {\tilde \Phi}_1 d_4 
- Y^u_{\beta 4} \overline{u_{\mathrm{R}}}_\beta \Phi_2^\dagger q_4 
- Y^d_{\beta 4} \overline{d_{\mathrm{R}}}_\beta {\tilde \Phi}_1^\dagger q_4 
\nonumber \\&&
- Y^\ell_{\alpha 4} \overline{\ell}_\alpha {\tilde \Phi}_1 e_4 
- Y^e_{\beta 4} \overline{e_{\mathrm{R}}}_\beta {\tilde \Phi}_1^\dagger \ell_4
- y^q_{\alpha 4} \phi\, \overline{q_L}_\alpha q_4 
- y^u_{\beta 4} \overline{\phi}\, \overline{u_{\mathrm{R}}}_\beta u_4 
\nonumber \\&&
- y^d_{\beta 4} \overline{\phi}\, \overline{d_{\mathrm{R}}}_\beta d_4 
- y^\ell_{\alpha 4} \phi\, \overline{\ell_L}_\alpha \ell_4  
- y^e_{\beta 4} \overline{\phi}\, \overline{e_{\mathrm{R}}}_\beta e_4 
 + {\rm h.c.}\,.
\label{eq:Yuk4f} 
\end{eqnarray}
The mass terms of the fourth family fermions are also imposed as
\begin{eqnarray}
\mathcal{L}_{\rm mass} & \supset & 
M^q_4 \overline{q_4} q_4 + M^u_{4} \overline{u_4} u_4 + M^d_{4} \overline{d_4} d_4 + M^\ell_{4} \overline{\ell_4} \ell_4 + M^e_{4} \overline{e_4} e_4 \,,
\label{eq:mass:quark} 
\end{eqnarray}
where all the masses of the vector-like fermions are considered to be far larger than the EW scale. Similar to the Weinberg operator in the seesaw mechanism, some dimension-5 effective operators can be generated by integrating out the fourth family fermion fields 
\begin{eqnarray}
\mathcal{L}_{\rm eff} &= &
- \frac{1}{M^u_4}Y^{qu}_{\alpha 4} (y^u_{\beta 4})^* \overline{q_L}_\alpha \Phi_2 \phi\, u_{\mathrm{R}\beta} 
- \frac{1}{M^q_4}y^q_{\alpha 4} (Y^u_{\beta 4})^* \overline{q_L}_\alpha \Phi_2 \phi\, u_{\mathrm{R}\beta}\nonumber\\&&
- \frac{1}{M^d_4}Y^{qd}_{\alpha 4} (y^d_{\beta 4})^* \overline{q_L}_\alpha {\tilde \Phi}_1 \phi\, d_{\mathrm{R}\beta} 
- \frac{1}{M^q_4}y^q_{\alpha 4} (Y^d_{\beta 4})^* \overline{q_L}_\alpha {\tilde \Phi}_1 \phi\, d_{\mathrm{R}\beta}\nonumber\\&&
- \frac{1}{M^e_4}Y^\ell_{\alpha 4} (y^u_{\beta 4})^* \overline{\ell}_\alpha {\tilde \Phi}_1 \phi\, e_{\mathrm{R}\beta} 
- \frac{1}{M^\ell_4}y^\ell_{\alpha 4} (Y^u_{\beta 4})^* \overline{\ell}_\alpha {\tilde \Phi}_1 \phi\, e_{\mathrm{R}\beta} + {\rm h.c.}
\label{eq:Yuk_eff} 
\end{eqnarray}
The diagrams for interaction between quark doublets and up-type quarks are shown in Fig.\ref{fig:u-quark} as an example.
After the Yukon $\phi$ gains a VEV, the resulting interactions are coincident with the Lagrangian in Eq.\eqref{eq:Yuk} with Yukawa couplings 
\begin{eqnarray}
Y^u_{\alpha \beta} = \frac{Y^{qu}_{\alpha 4} (y^u_{\beta 4})^*\langle\phi\rangle}{M^u_4} + \frac{y^q_{\alpha 4} (Y^u_{\beta 4})^*\langle\phi\rangle}{M^q_4} \,, \nonumber\\ 
Y^d_{\alpha \beta} = \frac{Y^{qd}_{\alpha 4} (y^d_{\beta 4})^*\langle\phi\rangle}{M^d_4} + \frac{y^q_{\alpha 4} (Y^d_{\beta 4})^*\langle\phi\rangle}{M^q_4} \,, \nonumber\\
Y^e_{\alpha \beta} = \frac{Y^\ell_{\alpha 4} (y^u_{\beta 4})^*\langle\phi\rangle}{M^e_4} + \frac{y^\ell_{\alpha 4} (Y^u_{\beta 4})^*\langle\phi\rangle}{M^\ell_4}  \,.
\label{eq:Yuk_couplings} 
\end{eqnarray}

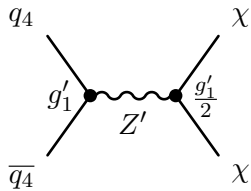
\begin{figure}[t!]
\begin{center}
\begin{fmffile}{FeynDiag/DMpro3}
\fmfframe(20,20)(20,20){
\begin{fmfgraph*}(80,45)
\fmflabel{$\chi$}{o1}
\fmflabel{$\chi$}{o2}
\fmflabel{$q_4$}{i2}
\fmflabel{$\overline{q_4}$}{i1}
\fmfv{label=$g_1'$}{v1}
\fmfv{label=$\frac{g_1'}{2}$}{v2}
\fmfleft{i1,i2}
\fmfright{o1,o2}
\fmf{plain}{i2,v1}
\fmf{plain}{i1,v1}
\fmf{plain}{v2,o1}
\fmf{plain}{v2,o2}
\fmf{photon,label=$Z'$}{v1,v2}
\fmfdotn{v}{2}
\end{fmfgraph*}}
\end{fmffile}
\end{center}
\caption{\label{fig:Feyn_q4} The process responsible for DM production in the complete model in addition to those in Fig.\ref{fig:Feyn}}
\end{figure}

Although the only constraint on the couplings in Eq.\eqref{eq:Yuk4f} is from the SM fermion mass matrix in a general basis, there exist a particular basis where some of the couplings can be zero \cite{King:2018fcg}. By rechoosing the chiral quark basis, we can find a basis where
\begin{eqnarray}
y^q_{\alpha 4} = \begin{pmatrix} 0 & 0  & y^q_{34} \end{pmatrix} \,, \quad
Y^{qu}_{\alpha 4} =&& \begin{pmatrix} 0 & Y^{qu}_{24}  & Y^{qu}_{34} \end{pmatrix} \,, \quad
Y^{qd}_{\alpha 4} = \begin{pmatrix} Y^{qd}_{14} & Y^{qd}_{24}  & Y^{qd}_{34} \end{pmatrix} \,, \quad \\
Y^u_{\beta 4} = \begin{pmatrix} 0 & 0  & Y^u_{34} \end{pmatrix} &&\,, \quad
y^u_{\beta 4} = \begin{pmatrix} 0 & y^u_{24} & y^u_{34} \end{pmatrix} \,, \\
Y^d_{\beta 4} = \begin{pmatrix} 0 & 0  & Y^d_{34} \end{pmatrix} &&\,, \quad
y^d_{\beta 4} = \begin{pmatrix} 0 & y^d_{24} & y^d_{34} \end{pmatrix} \,.
\end{eqnarray}
In such a basis, the quark couplings in Eq.\eqref{eq:Yuk_couplings} read
\begin{eqnarray}
Y^u_{\alpha \beta} &=&
\begin{pmatrix} 
0 & 0 & 0 \\ 
0 & Y^{qu}_{24} (y^u_{24})^* & Y^{qu}_{24}(y^u_{34})^* \\ 
0 & Y^{qu}_{34} (y^u_{24})^* & Y^{qu}_{34}(y^u_{34})^* 
\end{pmatrix}
\frac{\langle\phi\rangle}{M^u_4} + 
\begin{pmatrix} 
0 & 0 & 0 \\ 
0 & 0 & 0 \\ 
0 & 0 & y^q_{34}(Y^u_{34})^* 
\end{pmatrix}
\frac{\langle\phi\rangle}{M^q_4} \,, \\ 
Y^d_{\alpha \beta} &=& 
\begin{pmatrix} 
0 & Y^{qd}_{14} (y^d_{24})^* & Y^{qd}_{14}(y^d_{34})^* \\ 
0 & Y^{qd}_{24} (y^d_{24})^* & Y^{qd}_{24}(y^d_{34})^* \\ 
0 & Y^{qd}_{34}(y^d_{24})^* & Y^{qd}_{34}(y^d_{34})^* 
\end{pmatrix}
\frac{\langle\phi\rangle}{M^d_4} + 
\begin{pmatrix} 
0 & 0 & 0 \\ 
0 & 0 & 0 \\ 
0 & 0 & y^q_{34}(Y^d_{34})^* 
\end{pmatrix}
\frac{\langle\phi\rangle}{M^q_4} \,.
\end{eqnarray}
Without further modification, the first family quark remains massless in this basis. The problem can be solved by considering some more massive particles, such as a neutral Higgs messenger \cite{Ferretti:2006df}, which also helps to explain the quark mass hierarchy. Nevertheless, the massive particles are not likely to make any phenomenological effects other than the up quark mass and therefore are not going to be discussed further. To explain the heaviness of top quark, the mass of $q_4$ is assumed to be the lightest in the vector-like fermions. As the Yukawa coupling of top quark, namely $Y^u_{33}$, is determined by $m_{\rm t}/\langle\Phi_2\rangle\simeq1$, $M^q_4$ cannot be much larger than $\langle\phi\rangle$. On the other hand, the experimental limit from vector-like top (VLT) decay on $M^q_4$ is around 1 TeV \cite{King:2020mau}. If $q_4$ is not too heavy, it makes an extra contribution to the amplitude during the freeze-out in addition to the processes in Fig.\ref{fig:Feyn}, as shown in Fig.\ref{fig:Feyn_q4}. 
It can also affect the decay rate of the new gauge boson if $q_4$ is more than twice lighter than $Z'$.

Here, to quantify the influence of the vector-like fermions, we consider $q_4$ with a mass of 1 TeV and show the allowed parameter space in Fig.\ref{fig:ra_4th} in a similar style to the one in Fig.\ref{fig:ra_eff} for the effective model. Compared with the result in the effective model, the maximal masses of the dark fermion $\chi$ and the gauge boson $Z'$ increase to 31 TeV and 52 TeV respectively. As the allowed parameter space is enlarged, the perturbativity limit of the Yukawa-type interaction coupling in the dark sector starts to play a role in constraining the parameters. For TeV scale dark matter candidate, again, the relation $M_{Z'}\simeq2m_\chi$ is required by the constraint on $Z-Z'$ mixing angle.

\section{Conclusion \label{sec:Concl}}

In this paper, we have considered the possibility that dark matter is stabilised by a discrete $Z_2$ symmetry which arises from a subgroup of a 
$U(1)'$ gauge symmetry, spontaneously broken by integer charged scalars, and 
under which the chiral quarks and leptons do not carry any charges.
A four-component fermion $\chi$ with half-integer charge is odd under the preserved $Z_2$, and hence becomes a stable dark matter candidate, being produced through couplings to right-handed neutrinos with vector-like $U(1)'$ charges. For simplicity we have assumed that the lightest component of the four-component fermion is predominantly right-handed, $\chi_R$, although the results may be readily generalised to other cases as we have indicated.

We have constructed an effective model along these lines as an extension of the type Ib seesaw model where the light neutrino mass originates from a new type of Weinberg operator involving two Higgs doublets and a heavy Dirac neutrino. In such a model, the Majorana mass of the heavy neutrinos is forbidden by the $U(1)'$ gauge symmetry and therefore the usual type Ia seesaw mechanism is not allowed. Although the SM charged fermions cannot interact with the Higgs doublets through Yukawa type interaction due to the $U(1)'$ symmetry, they can gain mass from dimension five effective operators involving a Higgs singlet named Yukon, which is integer charged under the $U(1)'$ symmetry, and is responsible for breaking it to $Z_2$.
After the $U(1)'$ symmetry breaking, the dark matter candidate $\chi$ can only interact with the thermal bath through processes mediated by the $U(1)'$ gauge boson and therefore can be produced thermally in the early universe. 
 
We have explored the allowed parameter space of the effective model providing the correct dark matter relic abundance. Through analytical computation, we have found that the dark matter can only be produced correctly when there is no hierarchy between the masses of the dark fermion $\chi$ and the $U(1)'$ gauge boson. The numerical results show that there exists a resonance in the cross section when the dark fermion mass is half of the $U(1)'$ gauge boson. For this reason, the experimental bound on the massive gauge boson mixing prefers the line $M_{Z'}=2m_\chi$ for TeV scale dark matter candidate. Although the parameter space is not constrained by current experiments, we have estimated the required sensitivity for direct and indirect detections in this model.

We then considered a high energy renormalisable model with a complete fourth family of vector-like fermions, where the chiral quark and lepton masses arise from a seesaw-like mechanism. With the inclusion of the fourth family, the lightest vector-like quark can contribute to the dark matter production, enlarging the allowed parameter space that we explore. 
By integrating out the vector-like fermions, the non-renormalisable type Ib seesaw model can be obtained effectively with the charged fermion masses generated as in a type II 2HDM. Taking the contribution from the lightest fourth family quark into consideration, we have found that the allowed parameter space is enlarged, while the constraint on $Z-Z'$ mixing still keeps the relation $M_{Z'}=2m_\chi$ when the dark matter candidate is around TeV scale. 

In conclusion, we have proposed and explored a model which can account for both dark matter and neutrino mass and mixing, without requiring the addition of discrete symmetries to stabilise the dark matter mass. We have focussed on a fermiophobic $U(1)'$ model in which vector-like right-handed neutrinos form a Dirac neutrino mass and act as a portal for dark matter production, while at the same time providing a low scale testable seesaw mechanism referred to as type Ib since it involves two different Higgs doublets.

\section*{Acknowledgments}
We would like to thank Simon King for suggesting the mechanism studied here.
BF acknowledges the Chinese Scholarship Council (CSC) Grant No.\ 201809210011 under agreements [2018]3101 and [2019]536. SFK acknowledges the STFC Consolidated Grant ST/L000296/1 and the European Union's Horizon 2020 Research and Innovation programme under Marie Sklodowska-Curie grant agreement HIDDeN European ITN project (H2020-MSCA-ITN-2019//860881-HIDDeN).

\bibliographystyle{JHEP}
\bibliography{Type1b}

\end{document}